\documentclass[article,twocolumn,
superscriptaddress,
amsmath,amssymb,
aps,
prl,
floatfix,
]{revtex4-2}

\usepackage{float}
\usepackage{graphicx}
\usepackage{dcolumn}
\usepackage{bm}
\usepackage{hyperref}
\hypersetup{
    colorlinks=true,
    linkcolor=blue,
    filecolor=magenta,      
    urlcolor=cyan,
}
\usepackage{color}
\usepackage{fancyhdr}
\usepackage[normalem]{ulem}

\newcommand{\risp}[1]{\textcolor{black}{{#1}}}

\usepackage{tikz,xcolor,hyperref}

\definecolor{lime}{HTML}{A6CE39}
\DeclareRobustCommand{\orcidicon}{
	\begin{tikzpicture}
	\draw[lime, fill=lime] (0,0) 
	circle [radius=0.16] 
	node[white] {{\fontfamily{qag}\selectfont \tiny ID}};
	\draw[white, fill=white] (-0.0625,0.095) 
	circle [radius=0.007];
	\end{tikzpicture}
	\hspace{-2mm}
}
\foreach \x in {A, ..., Z}{\expandafter\xdef\csname orcid\x\endcsname{\noexpand\href{https://orcid.org/\csname orcidauthor\x\endcsname}
			{\noexpand\orcidicon}}
}

\begin{document}

\title{Coulomb crystallization of xenon highly charged ions in a laser-cooled Ca$^{+}$ matrix}

\author{Leonid Prokhorov\orcidB{}}
\affiliation{School of Physics and Astronomy, University of Birmingham, Edgbaston, Birmingham, B15 2TT, United Kingdom}
\author{Aaron A. Smith}
\affiliation{School of Physics and Astronomy, University of Birmingham, Edgbaston, Birmingham, B15 2TT, United Kingdom}
\author{Mingyao Xu}
\affiliation{School of Physics and Astronomy, University of Birmingham, Edgbaston, Birmingham, B15 2TT, United Kingdom}
\author{Kostas Georgiou}
\affiliation{School of Physics and Astronomy, University of Birmingham, Edgbaston, Birmingham, B15 2TT, United Kingdom}
\affiliation{Max-Planck-Institut f\"ur Kernphysik, D--69117 Heidelberg, Germany}
\author{Vera Guarrera\orcidC{}}
\affiliation{School of Physics and Astronomy, University of Birmingham, Edgbaston, Birmingham, B15 2TT, United Kingdom}
\author{Lakshmi P. Kozhiparambil Sajith}
\affiliation{Max-Planck-Institut f\"ur Kernphysik, D--69117 Heidelberg, Germany}
\affiliation{Institut f\"ur Physik, Humboldt-Universit\"at zu Berlin, Newtonstra{\ss}e 15, 12489 Berlin, Germany}                                    
\affiliation{Deutsches Elektronen-Synchrotron (DESY), Platanenallee 6, D-15738 Zeuthen, Germany}    
\author{Elwin A. Dijck\orcidF{}}
\affiliation{Max-Planck-Institut f\"ur Kernphysik, D--69117 Heidelberg, Germany}
\author{Christian Warnecke\orcidE{}}
\affiliation{Max-Planck-Institut f\"ur Kernphysik, D--69117 Heidelberg, Germany}
\affiliation{Institut f\"ur Physik, Humboldt-Universit\"at zu Berlin, Newtonstra{\ss}e 15, 12489 Berlin, Germany}                                    
\affiliation{Deutsches Elektronen-Synchrotron (DESY), Platanenallee 6, D-15738 Zeuthen, Germany}
\author{Malte Wehrheim\orcidM{}} 
\affiliation{Physikalisch-Technische Bundesanstalt, 38116 Braunschweig, Germany}
\author{Alexander Wilzewski} 
\altaffiliation{current address: Max-Planck-Institute for Quantum Optics, 85748 Garching, Germany}
\affiliation{Physikalisch-Technische Bundesanstalt, 38116 Braunschweig, Germany}
\author{Laura Blackburn\orcidM{}} 
\affiliation{Department of Physics and Astronomy, University of Sussex, Brighton, BN1 9QH, United Kingdom}
\author{Matthias Keller\orcidH{}} 
\affiliation{Department of Physics and Astronomy, University of Sussex, Brighton, BN1 9QH, United Kingdom}
\author{Vincent Boyer\orcidI{}}
\affiliation{School of Physics and Astronomy, University of Birmingham, Edgbaston, Birmingham, B15 2TT, United Kingdom}
\author{Thomas Pfeifer\orcidL{}}
\affiliation{Max-Planck-Institut f\"ur Kernphysik, D--69117 Heidelberg, Germany}
\author{Ullrich Schwanke\orcidP{}}
\affiliation{Institut f\"ur Physik, Humboldt-Universit\"at zu Berlin, Newtonstra{\ss}e 15, 12489 Berlin, Germany}   
\author{Cigdem Issever\orcidO{}}
\affiliation{Institut f\"ur Physik, Humboldt-Universit\"at zu Berlin, Newtonstra{\ss}e 15, 12489 Berlin, Germany}                                    
\affiliation{Deutsches Elektronen-Synchrotron (DESY), Platanenallee 6, D-15738 Zeuthen, Germany}  
\author{Steven Worm\orcidD{}}
\affiliation{Institut f\"ur Physik, Humboldt-Universit\"at zu Berlin, Newtonstra{\ss}e 15, 12489 Berlin, Germany}                                    
\affiliation{Deutsches Elektronen-Synchrotron (DESY), Platanenallee 6, D-15738 Zeuthen, Germany}    
\author{Piet O. Schmidt\orcidG{}} 
\affiliation{Physikalisch-Technische Bundesanstalt, 38116 Braunschweig, Germany}
\affiliation{Institute for Quantum Optics, Leibniz University Hannover, 30167 Hannover, Germany}
\author{Jos\'e R. {Crespo L\'opez-Urrutia}\orcidJ{}}
\affiliation{Max-Planck-Institut f\"ur Kernphysik, D--69117 Heidelberg, Germany}
\author{Giovanni Barontini \orcidA{}}
\email{g.barontini@bham.ac.uk}
\affiliation{School of Physics and Astronomy, University of Birmingham, Edgbaston, Birmingham, B15 2TT, United Kingdom}

\date{\today}

\begin{abstract}
We report on the sympathetic cooling and Coulomb crystallization of xenon highly charged ions (HCIs) with laser-cooled Ca$^+$ ions. The HCIs are produced in a compact electron beam ion trap, then charge selected, decelerated, and finally injected into a cryogenic linear Paul trap. There, they are captured into $^{40}$Ca$^+$ Coulomb crystals, and co-crystallized within them, causing dark voids in their fluorescence images. Fine control over the number of trapped ions and HCIs allows us to realize mixed-species crystals with arbitrary ordering patterns. By investigating Xe$^{q+}$--Ca$^+$ strings, we confirm the HCI charge states, measure their lifetime and characterize the mixed-species motional modes. Our system effectively combines the established quantum control toolbox for Ca$^+$ with the rich set of atomic properties of Xe highly charged ions, providing a resourceful platform for optical frequency metrology, searches for signatures of new physics, and quantum information science.
\end{abstract}

\maketitle
Highly charged ions (HCIs) provide a promising new platform for next generation atomic clocks, precision spectroscopy, and searches for physics beyond the Standard Model \cite{Safronova2018RMP}. The strong Coulomb binding of the outermost electrons in HCIs makes them relatively insensitive to external fields. Fine-structure, narrow transitions of these electrons over a broad spectral range \cite{Kozlov2018RMP,Dzuba2015HCI,schiller_hydrogenlike_2007,berengut_enhanced_2010} provide ideal systems to implement accurate atomic clocks. HCIs have indeed the potential to extend optical clock operations into the deep ultraviolet and extreme-ultraviolet spectral regions \cite{Kozlov2018RMP} that could represent the next step in atomic clock technology. Furthermore, due to the compact wavefunctions of the outer electrons that enhance relativistic effects, specific HCIs have clock transitions with an amplified response to changes of the fine structure constant $\alpha$ \cite{Kozlov2018RMP,Dzuba2015HCI,berengut_enhanced_2010,Berengut2012PRL,Porsev2020PRA,barontini2022measuring}, or to violations of Lorentz invariance \cite{Shaniv2018PRL}. Clock comparisons based on these systems are predicted to have sensitivities to new physics one to two orders of magnitude higher than those of existing neutral atom and singly charged ion clocks \cite{Safronova2018RMP,Sherrill2023NJPhys,Beloy2021Nature,Filzinger2023PRL}. 

On the experimental side, \risp{pioneering Penning trap experiments demonstrated sympathetic cooling of highly charged xenon ions in Be$^+$ Coulomb crystals \cite{gruPhysRevLett.86.636,hold10.1063/1.1454280}, reaching temperatures of the order of 1~K. Building on this foundation, the last decade has seen rapid progress toward bringing trapped HCIs into regimes suitable for precision spectroscopy and metrology. In particular, sympathetic cooling of Ar$^{13+}$ in Be$^+$ Coulomb crystals in a Paul trap demonstrated that these systems can be prepared in the mK regime \cite{Schmoeger2015Science}.} Subsequent advances, including quantum logic spectroscopy \cite{Micke2020Nature}, the first optical atomic clock \cite{King2022Nature}, and a King plot analysis of the isotope shift of the fine structure in Ca$^{14+}$ to constrain fifth forces \cite{Wilzewski2025}, have established HCIs as a realistic platform for metrology \cite{Spiess2025}, quantum control, and tests of fundamental physics \cite{Wilzewski2025}. Coulomb crystallization of Ni$^{12+}$ in a Be$^+$ matrix and the observation of its clock transition have extended these techniques to another promising atomic clock candidate \cite{Chen2024,Chen2025PRL,ChenShaolong2025}.

In parallel, theoretical studies and spectroscopic measurements have identified xenon HCIs as especially attractive for the implementation of optical atomic clocks and for probing new physics. In particular, a recent comprehensive spectroscopic study of Xe$^{q+}$ found a series of narrow and ultranarrow transitions, which are promising clock candidates \cite{Rehbehn2023PRL}. In addition, Xe offers several stable isotopes, making it ideally suited for isotope shift measurements and generalized King plot analyses \cite{Kozlov2018RMP, Rehbehn2023PRL,Wilzewski2025, berengut_generalized_2020}. 

Realizing the full potential of Xe HCIs for quantum technologies requires the ability to sympathetically cool them within a host Coulomb crystal. Coulomb crystals of laser-cooled Ca$^+$ ions are a workhorse platform in trapped ion quantum science: they combine a simple level structure with a narrow clock transition, and mature control techniques developed for optical frequency standards and quantum information processing \cite{Huang2011PRA,zhang_liquid-nitrogen-cooled_2025,marciniak_optimal_2022,pogorelov_compact_2021}.   Ca$^+$ Coulomb crystals have been extensively used as host matrices for sympathetically cooled ``dark'' impurity ions, including molecular ions and other atomic species \cite{PhysRevLettxx.105.143001,chou2017preparation,krohn2023reactions,Drewsen2015PhysicaB,Blackburn2020SciRep,Blackburn2025NJP,khanyile2015observation,rugango2015sympathetic,okPhysRevApplied.4.054009, petralia2020strong,guggemos2015sympathetic}. In addition, as $^{40}$Ca$^+$ has a comparatively low charge-to-mass ratio, Coulomb crystals of this species are suited for embedding HCIs of heavy species, such as those that feature enhanced sensitivity to variations of fundamental constants, e.g., Cf$^{15+}$ and Cf$^{17+}$ \cite{Berengut2012PRL,Porsev2020PRA,barontini2022measuring}.

In this Letter we demonstrate Coulomb crystallization of Xe HCIs in a laser-cooled Ca$^+$ matrix trapped in a linear Paul trap. Starting from Xe$^{q+}$ bunches extracted from a compact electron beam ion trap (EBIT), we decelerate and retrap the HCIs into a preformed $^{40}$Ca$^+$ Coulomb crystal, where they manifest themselves as well localized ``dark'' impurities. We then demonstrate single-atom control on our mixed-species Coulomb crystals, that enables the engineering of arbitrary configurations. We finally characterize our mixed-species linear crystals, in particular by measuring their normal-mode structure. Our results set the stage for future Xe-based HCI optical clocks and precision tests of fundamental physics.

\begin{figure*}
\centering
\includegraphics[width=\textwidth]{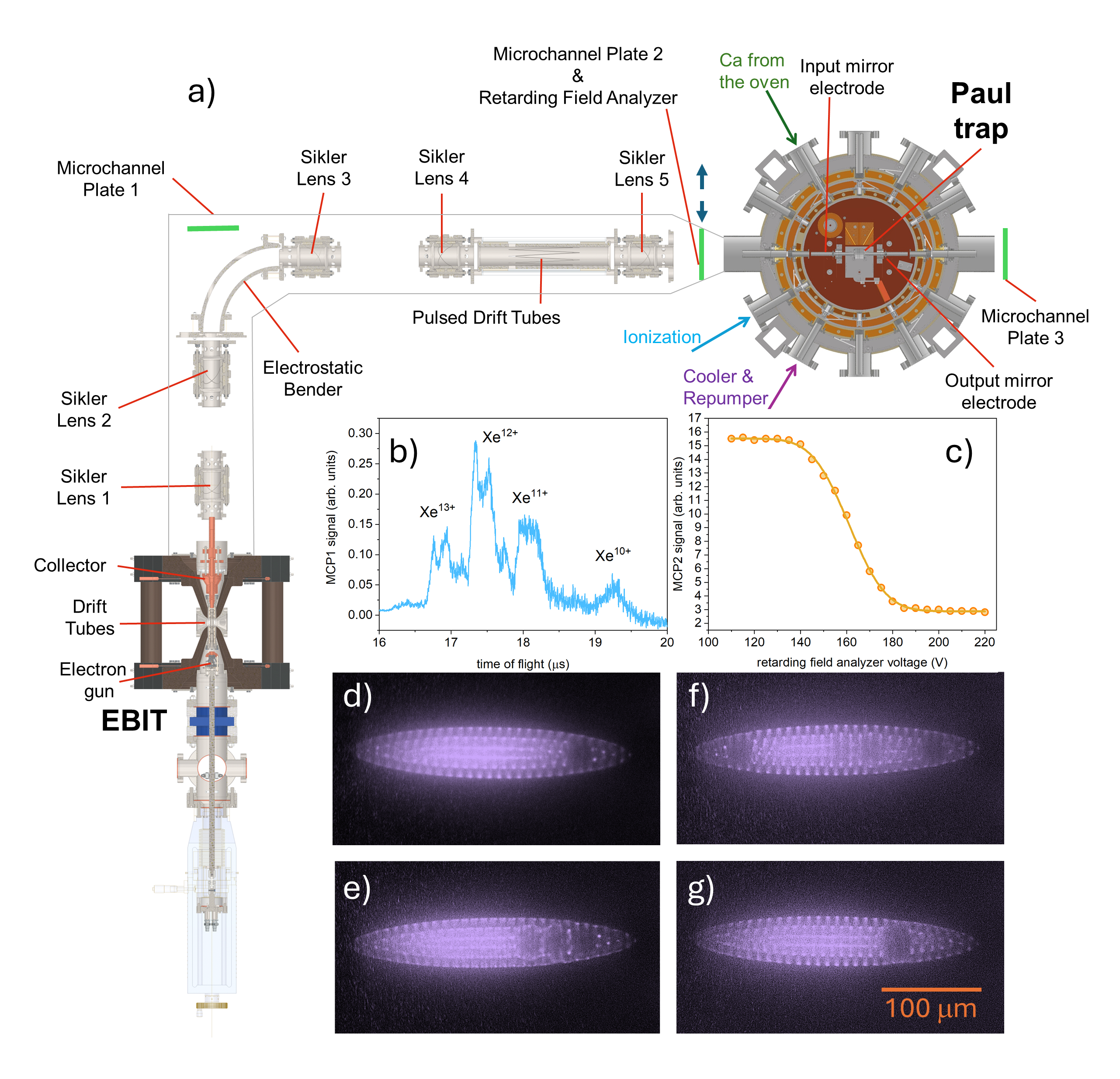}
\caption{a) Schematics of the main in-vacuo components of our apparatus, including the compact EBIT, the beam line, and the science chamber (relative distances are not to scale). The science chamber has concentric room temperature, 40\,K, and 4\,K stages, with the cryogenic Paul trap at the center. Arrows show the ionization and cooling lasers, and the Ca atomic beam injection into the cryogenic region. b) Typical time-of-flight spectrum of Xe HCIs as observed on microchannel plate 1. Large peaks correspond to the labelled ionization states, and smaller features to different stable isotopes of Xe. c) Points display the amplitude of the signal of charge-selected Xe$^{11+}$ ions on microchannel plate 2 as a function of the voltage of the retarding field analyser, with a fitted line using an error function. This fit yields a central energy of $\simeq$161 $q$\,eV, where $q$ is the charge state, and a full-width-at-half-maximum of $\simeq$12.6 $q$\,eV. This microchannel plate can be moved out to allow the HCIs to reach the science chamber. d-f) Fluorescence images of Ca$^+$ Coulomb crystals with 1, 3 and 4 Xe$^{11+}$ HCI implanted, respectively. g) the same as d) but with one Xe$^{19+}$ HCI crystallized. All exposure times are 1\,s. For single Ca$^+$ ion we measure trapping frequencies of $(\omega_x,\omega_y,\omega_z)=2\pi\times(821, 844,234)$\,kHz, respectively.}
\label{Fig1}
\end{figure*}

Our experimental system is based on those in \cite{Schwarz2012CryoPaul,schmo10.1063/1.4934245, Micke2018HC_EBIT, Micke2019Cryostat,leopold_cryogenic_2019}. In essence, it combines an EBIT \cite{Micke2018HC_EBIT} to produce HCIs on one end, with an extremely high vacuum cryogenic environment to trap and cool the HCIs \cite{Micke2019Cryostat,leopold_cryogenic_2019} on the other end, and a beam line that guides the HCIs in between. In Fig. \ref{Fig1} a) we show the main components of our EBIT and beam line, which have been described in detail in \cite{Micke2018HC_EBIT, Dijck2023RSI}.  Cryogenic cooling for the science chamber [(top right of Fig. \ref{Fig1} a)] is provided by a pulse tube cryocooler (not shown in the figure), with nominal capacities of 1\,W at 4.2\,K and 40\,W  at 45\,K. To ensure good thermal connection between the cryocooler and the science chamber, together with substantial vibration isolation, we use a thermal transfer unit as the one described in \cite{Micke2019Cryostat} (not shown). At the center of the science chamber, inside the 4 K stage, there is a cryogenic linear segmented Paul trap, which is a variant of the one described in \cite{leopold_cryogenic_2019}. Its four blades have a separation of 0.7\,mm. Two of the blades are connected to an RF resonator, also hosted inside the 4\,K stage, to provide radial confinement. The other two blades are split into 5 sections and are used to provide axial confinement.  All the DC voltages are applied to the trap through a PCB board hosting low-pass filters with a cut-off frequency at 200\,Hz. The PCB is located inside the 4 K stage to minimize Johnson noise and to ensure that there is no heat load from the wires to the trap electrodes. The filter board and RF share the same ground, which can be elevated to hundreds of V for HCI deceleration, as explained later. 

Unlike other systems, our experiment utilizes Coulomb crystals of Ca$^+$ ions for sympathetic cooling of the HCIs. \risp{Neutral Ca atoms are introduced into the trapping region from a dedicated resistively heated oven, as shown in Fig.~\ref{Fig1}(a). $^{40}$Ca$^+$ ions are selectively produced within the trapping region by resonant two-step photoionization using light at 423~nm and 375~nm. Doppler cooling is performed on the Ca$^+$ cycling transition using a cooling beam at 397~nm overlapped with an 866~nm repumper. All optical sources are diode lasers, with their frequencies stabilized using a wavelength meter.} The fluorescence of trapped Ca$^+$ ions is imaged with a system that comprises an in-vacuum high numerical aperture aspheric lens that produces a 3$\times$ magnified image outside the vacuum system \cite{leopold_cryogenic_2019}, and a microscope assembly that further magnifies this latter image $10\times$ on an EMCCD camera (Andor iXon Ultra 888).   

Our experimental sequence broadly traces the ones described in \cite{Schmoeger2015Science,Dijck2023RSI}. In this work, the Xe HCIs are extracted from the EBIT every 250 ms, with the extraction energy depending on the ionization state. To provide a concrete example, in the following we describe the typical sequence for Xe$^{11+}$ HCIs. For this ionization state we use an electron-beam energy of $\simeq$ 280 eV, while the extraction energy is 400 $q$eV, with $q=11$. After extraction, the HCIs are directed and focused using a series of Sikler lenses \cite{Sikler2005,SiklerMandal2011}, as shown in Fig. \ref{Fig1} a). We utilise the signal of microchannel plate 1 to maximise the number of HCIs extracted. An example of a time-of-flight spectrum obtained on this microchannel plate is shown in Fig. \ref{Fig1} b). The peaks of the different ionization states are rather broad, have internal structure, and partially overlap due to the fact that Xe has a large number of stable isotopes. \risp{Note that the shape of the peaks and their substructure depend on the extraction settings, namely the extraction voltage and the voltages of the first two Sikler lenses.}

After the electrostatic bender, the HCIs are charge-selected using Sikler lens 3 as a timed gate. They are then decelerated and bunched in energy using a pulsed-drift tube whose first electrode is kept at 250\,V, while the second is at 600 V. Both potentials are switched to zero while the HCIs are traveling through the tube. The second microchannel plate is placed after the pulsed-drift tube and has incorporated a retarding-field analyzer that is used to measure the energy distribution of the HCI bunch, as shown in Fig. \ref{Fig1} c). With the above settings, we typically obtain decelerated bunches with mean energies of $\simeq$ 160\,$q$eV and full width at half maximum of $\simeq$ 10-15\, $q$eV. The spread in energy is substantially larger than in Ar HCIs (see, e.~g., \cite{Dijck2023RSI}), and we attribute this to the overlapping signals coming from different isotopes.  

The HCIs finally enter the science chamber [see Fig. \ref{Fig1} a)] and pass through the input mirror electrode, which is kept at 0\,V \risp{\footnote{\risp{No active isotope separation is implemented prior to trap injection in the present work; the loaded ions therefore reflect the natural isotopic mixture.}}}. The Paul trap, which comes immediately after the input mirror electrode, is instead kept at $\simeq$145~V, in order to slow down the Xe HCIs to 1-20 $q$eV. The output mirror electrode is kept at 200 V and is used to reflect the HCIs back towards the center of the trap. Immediately after the HCIs exit the input mirror electrode, we raise its voltage to 200 V. In this way we force the HCIs to bounce back and forth between the two mirror electrodes, i.~e., we axially confine them inside the Paul trap. During this stage, we typically apply $\simeq$0.11 V to the DC electrodes and $\simeq$75 V to the RF electrodes of the Paul trap. This provides radial confinement for the HCIs.

We utilize the same Paul trap pseudopotential to confine Coulomb crystals of hundreds of Ca$^+$ ions. The HCIs that are bouncing back and forth between the mirror electrodes thus can interact several times and be sympathetically cooled by the laser cooled Ca$^+$ Coulomb crystal, until they finally crystallize. The presence of a HCI in the crystal manifests as a large round void in the Ca$^+$ fluorescence images [Fig. \ref{Fig1} d)-g)]. Once one or more HCIs  are in the crystal, we typically raise the RF voltage to $\simeq$200 V, as done in Fig. \ref{Fig1} d)-g). The probability of a HCI to eventually crystallize is roughly proportional to the number of Ca atoms in the crystal. We found that the probability also increases by using weaker pseudopotential confinement. In this work we mostly utilize Xe$^{11+}$ [Fig. \ref{Fig1} d)-f)]. By changing the parameters of the EBIT (electron beam energy) however, we are able to trap other ionization states. For example, in Fig. \ref{Fig1} g) we show a crystallized Xe$^{19+}$ ion which, to date, is the highest-charge ion ever crystallized.

\begin{figure}
\centering
\includegraphics[width=0.48\textwidth]{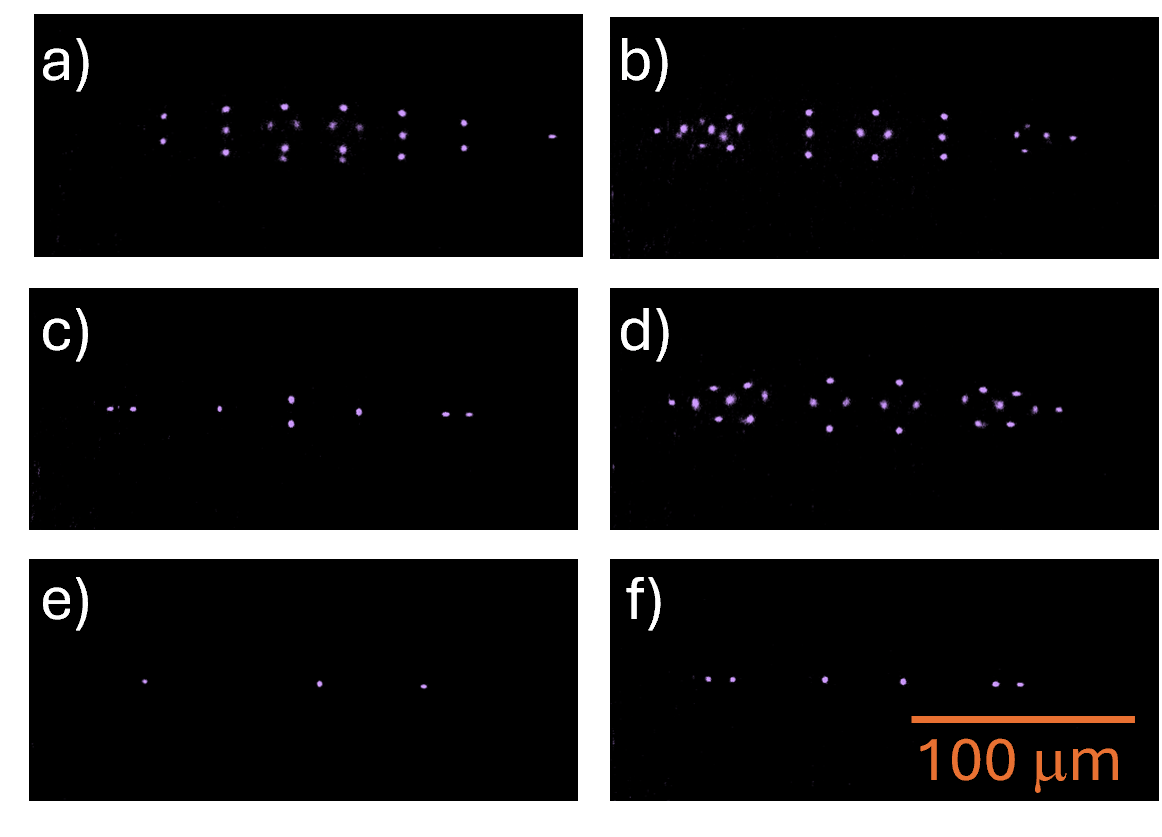}
\caption{a) Fluorescence images of a mixed-species Coulomb crystal containing seven Xe$^{11+}$ HCIs, each manifesting as a void in the crystalline structure. b) and c) Mixed-species Coulomb crystals with four Xe$^{11+}$ HCIs and different numbers of Ca$^+$ ions. d)-f) The same but with three Xe$^{11+}$ HCIs. The exposure time for all panels is 1 s.}
\label{Fig3}
\end{figure}

The number of ions in the Ca$^+$ Coulomb crystal can be controllably decreased by blue-detuning the cooling laser by $\simeq40$ MHz and reducing its intensity for a few seconds. This progressively expels ions from the trap. The time needed to expel one ion is approximately inversely proportional to the size of the crystal. The same procedure can be used when one HCI is co-trapped. The presence of the HCI speeds up the process for large crystals, while it has less effect for crystals of less than 10 atoms. When more than one HCI are trapped, the removal of Ca$^+$ atoms with this method becomes more delicate. In this case, to avoid sudden loss of the entire crystal during the blue-detuned phase, we must reduce the trapping frequencies to the point where the crystal is completely melted, and strongly reduce the power of the cooling laser. The number of Ca$^+$ ions is tracked via their total fluorescence. Once the desired number of ions is reached, we can restore stronger confinement to recrystallize. With this method we can engineer mixed Coulomb crystals with single atom precision. Some examples with up to seven HCIs are shown in Fig. \ref{Fig3}.

\begin{figure}
\centering
\includegraphics[width=0.48\textwidth]{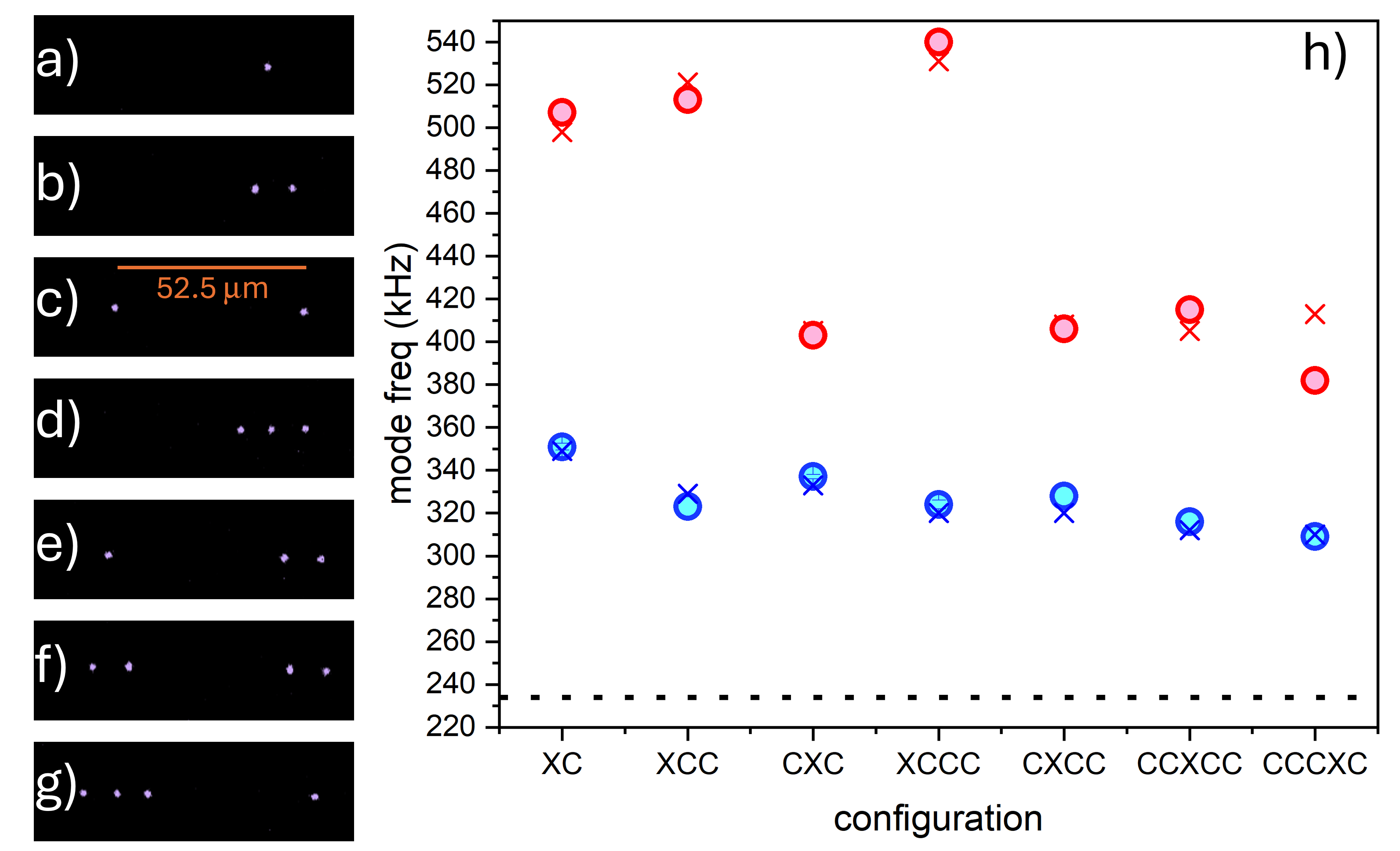}
\caption{a) Fluorescence image of a mixed-species Coulomb crystal containing one Ca$^+$ ion (the purple dot) and one Xe$^{11+}$ HCI (not visible). b-g) the same as a) but with two to four Ca$^+$ ions. h) The circles are the measured frequencies of the two lower axial modes of the crystals shown in a)-g), where X$\equiv$Xe$^{11+}$ and C$\equiv$Ca$^{+}$. The errorbars, \risp{derived from shot-to-shot fluctuations}, are smaller than the size of the circles. The crosses are the values calculated with the method explained in the text. The dashed line is the measured secular axial frequency for Ca$^+$ in the same trap.}
\label{Fig2}
\end{figure}

In Fig. \ref{Fig2} a)-g) we show a series of different configurations for small linear Xe$^{11+}$-Ca$^+$ Coulomb crystals with one HCI implanted, obtained using the method just described. Also in this case, the presence of a Xe HCI manifests as a dark void in the linear chain. For crystals that have the same number of atoms, different configurations (i.e. Xe$^{11+}$ in different positions) can be obtained by briefly modulating the amplitude of the RF at a frequency close to one of the trapping frequencies. These crystals are useful to run some simple diagnostics on the HCIs. For example the configuration of Fig. \ref{Fig2} c) can be used to measure their charge. The separation between the two Ca ions when a HCI is in between is given by:
\begin{equation}
    d=2\left[ \frac{e^2}{4\pi\epsilon_0}\frac{q+1/4}{m_{\text{Ca}}\omega_{\text{Ca}}^2} \right]^{1/3},
\end{equation}
with $m_{\text{Ca}}$ the mass and $\omega_{\text{Ca}}$ the axial secular frequency of the Ca$^+$ ions. For $q=11$ we have that $d\simeq52.5$ $\mu$m, in accordance with what is measured in the experiment. We also use the same configuration to measure the charge lifetime of the HCIs by monitoring the separation of the two Ca ions as a function of time. The average lifetime of a Xe$^{11+}$ HCI in our experiment is $\simeq27$ minutes. \risp{By assuming that ion losses are dominated by charge-exchange collisions with H$_2$ in the capture-limited (Langevin) regime}, we infer that the pressure inside the 4 K shield is $\simeq2\times10^{-14}$ mbar. As in other similar experiments, the lifetime drops significantly ($<$ 1 minute) after a few weeks. \risp{This is because residual gases (primarily H$_2$) gradually load and partially saturate the cryosorbing surfaces, while more weakly cryopumped species (e.g., He) accumulate in the chamber, reducing the effective cryo-pumping efficiency. A brief warm-up to $\simeq 20$\,K mobilizes these species, allowing them to be removed by the turbomolecular pumps and thereby restoring the cryopumping performance and the HCI lifetime.}

We finally characterize the linear mixed-species Coulomb crystals in Fig. \ref{Fig2} a)-g) by measuring their lowest normal axial modes. The modes of these crystals are indeed strongly influenced by the presence of the HCI, which can, e.~g., split the ensemble in domains \cite{Rueffert2024} that can be partially decoupled. To this end we modulate the amplitude of the RF drive of the Paul trap at different frequencies until we excite some axial motion of the crystal. This is detected by the broadening of the fluorescence signal of the single ions along the axial direction. The results are the round dots reported in Fig. \ref{Fig2} h). To evaluate the frequencies of the modes of the mixed-species crystals, we model the ions as point charges confined in a harmonic axial pseudopotential and interacting via the full pairwise Coulomb repulsion, and we restrict the solutions to the linear chain regime where all ions lie on the trap axis. We first determine the 1D equilibrium positions $z_i$ by solving the force-balance equations for all ions. Around this equilibrium, we expand the total potential $U$ to second order in small axial displacements, which yields the Hessian matrix of second derivatives $d^2U/dz_i dz_j$. Diagonalising the mass-weighted Hessian gives the axial normal modes \cite{HOME2013231}. The crosses in Fig. \ref{Fig2} h) are the two lower axial mode frequencies calculated with this method. They track the experimental values reasonably well and reproduce the same dependence on the crystal configuration, confirming the charge and mass assigned to the HCIs.  

In this work we have reported on a new system for the production, cooling and trapping of HCIs. We demonstrated the first use of laser cooled Ca$^+$ Coulomb crystals to sympathetically cool Xe HCIs. This led to the crystallization of Xe HCIs in the Ca$^+$ matrix. We have also implemented a method that enables us to accurately control the number of ions in the mixed-species Coulomb crystals. We have characterized our system and benchmarked it by measuring the axial normal modes of linear Xe$^{11+}$–Ca$^+$ linear crystals. Our results represent a step-change towards Xe-based HCI optical clocks and precision tests of fundamental physics. This is because of the  availability of several candidate clock transitions at accessible wavelengths that can also be used as probes of fifth forces \cite{Kozlov2018RMP,Rehbehn2023PRL,Wilzewski2025}. Our system also enables the combination of the quantum control toolbox developed for Ca$^+$ in quantum information and frequency metrology with the largely unexplored potential of cold HCIs. Techniques such as ground-state cooling, sideband spectroscopy, and quantum logic with Ca$^+$ can now be applied to Xe HCIs \cite{Drewsen2015PhysicaB,Huang2011PRA,Gao2020NSR,Micke2020Nature,King2022Nature,king_algorithmic_2021}. 
\risp{In this respect, a near-term priority is to identify the most promising clock-transition candidate in Xe HCIs and to deploy the spectroscopy laser systems required to address the targeted transition, including frequency stabilization to an ultra-stable reference. In parallel, we will implement quantum logic spectroscopy using the co-trapped Ca$^+$ ion for state preparation and readout. We also aim to develop deterministic isotope selection and loading capabilities, for example via enriched xenon gas sources and/or by incorporating mass-selective filtering prior to trapping.}
Because the Ca$^+$ clock transition is well developed as a high accuracy optical standard, a co-trapped Ca$^+$–Xe$^{q+}$ system can realize two optical clocks in the same trap. Comparing their frequencies within a single Coulomb crystal suppresses many common mode environmental and trap induced shifts, potentially providing a substantial advantage in clock-clock comparison experiments. Finally, advanced schemes to realize mixed-species entangling gates \cite{tan2015multi,Bruzewicz2019npjQI} could be adapted to Ca$^+$–Xe$^{q+}$, exploiting the narrow optical transitions and the common axial modes studied here. 

\paragraph*{Acknowledgements}
We acknowledge fruitful discussions with the members of the QSNET consortium, and David Lucas. We are grateful to Ted Forgan and Mingee Chung for their contribution in the early stages of the experiment, and to Xen Sergi, John Perrins, and Mark Wicks for the technical support. This work was supported by STFC and EPSRC under grants ST/Y00454X/1, ST/W006138/1, ST/T00603X/1, and ST/Y004418/1, by the European Partnership on Metrology, co-financed by the European Union’s Horizon Europe Research and Innovation Programme and by the Participating States, under grant number 23FUN03 HIOC; by Deutsche Forschungsgemeinschaft (DFG, German Research Foundation) under Germany’s Excellence Strategy—EXC-2123 QuantumFrontiers—390837967; and by the European Research Council (ERC) under the European Union’s Horizon 2020 research and innovation program (Grant Agreement No. 101019987, FunClocks). The project was supported by the Max-Planck Society, the Max-Planck–Riken–PTB–Center for Time, Constants and Fundamental Symmetries, the Helmholtz Association of German research Centres, and the Ministry of Science, Research and Culture of the State of Brandenburg within the Center for Quantum Technology and Applications (CQTA).  \begin{figure}[!h]
\centering
\includegraphics[width=0.1\textwidth]{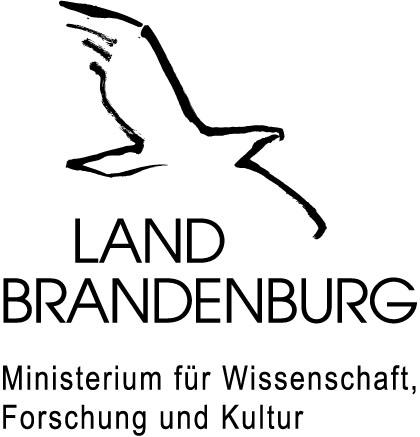}
\end{figure}   

\bibliography{main_bibl}

\end{document}